\documentclass[12pt,preprint]{aastex}
\shorttitle{Rapidly Spinning Black Holes in Quasars: An Open
Question} \shortauthors{Rafiee and Hall}
\begin{document}
\title{Rapidly
Spinning Black Holes in Quasars: An Open Question}
\author{Alireza Rafiee, Patrick B. Hall}
\affil{Department of Physics \& Astronomy, York University,
Toronto, Ontario M3J 1P3, Canada}
\begin{abstract}
Wang et al. (2006) estimated an average radiative efficiency of
30\%--35\% for quasars at moderate redshift.
We find that their method is not independent of quasar lifetimes
and thus that quasars do not necessarily have such high
efficiencies. Nonetheless, it is possible to derive interrelated
constraints on quasar lifetimes, Eddington ratios, and radiative
efficiencies of supermassive black holes. We derive such
constraints using a statistically complete sample of quasars with
black hole mass estimates from broad \ion{Mg}{2}, made both with
and without the radiation pressure correction of Marconi et al. (2008).
We conclude that for quasars with $L/L_{Edd}\gtrsim 0.02$, lifetimes
can range from 140 to 750~Myr for Schwarzschild black holes.
Coupled with observed black hole masses,
quasar lifetimes of $\leq$140~Myr would imply that radiatively inefficient accretion
or BH mergers must be important in the accretion history of quasars.
Given reasonable assumptions about the quasar population, if the average quasar lifetime
is $<1$ Gyr, and if radiatively inefficient accretion is unimportant, then not many BHs
with Eddington ratio $< 0.2$ can be rapidly spinning.

\end{abstract}

\section{Introduction}
A quasar is powered by matter accreting onto a supermassive black hole
(e.g., Rees 1984). Gas orbiting in the innermost stable circular
orbit (ISCO) may be perturbed and fall into the black hole (BH),
adding mass and angular momentum to it. To have reached the ISCO
{\em through a thin accretion disk}, gas must have radiated away a
fractional binding energy per unit rest mass $\simeq
1-(1-2GM/3c^2R_{ISCO})^{1/2}$ which is the system's
\textit{radiative efficiency} $\eta$ (Bardeen 1970). Since
$R_{ISCO}$ decreases from $6GM/c^2$ for Schwarzschild
(nonrotating) BHs to $GM/c^2$ for co-aligned accretion onto
Extreme-Kerr (maximally rotating) BHs, more energy is produced by
(co-aligned) thin disk accretion of a given mass onto a rotating
BH than onto a non-rotating BH (Carter 1968).

If a quasar accretes gas with a fixed angular momentum vector at
a fixed mass accretion rate $\dot{M}$, the BH spin will increase
to a theoretical maximum and the luminosity $L$ of the quasar will
increase along with it, since~$L\propto \eta \dot{M}c^2$.  In many
realistic accretion models, the spin of a supermassive BH
increases rapidly and then fluctuates around a maximum value,
although spin-down is also possible (Volonteri 2007).  Several such
models, including numerical ones by Di Matteo et al. (2005),
suggest that the accretion process stops when a BH becomes massive
enough to support a kinetic and/or radiative luminosity capable of
blowing away the gas fuelling it (Silk \& Rees 1998; Fabian 1999;
King 2003). However, the radiative luminosity for a given mass
accretion rate depends upon the radiative efficiency. A rapidly
rotating BH can help shut down the accretion process earlier than
a nonrotating BH could. Knowledge of supermassive BH spins is
therefore useful in constraining models of quasar development.

Observationally, the
quasi-periodic variability detected in Sgr A* may be evidence of
rapid spin of the Galactic BH (e.g., Genzel et
al. 2003). Evidence for rotating supermassive BHs comes from
studies of X-ray Fe~K$\alpha$ line profiles (Blandford et al.
1990) and by theoretical arguments that powerful radio jets are
powered by the extraction of energy from rotating BHs (Miller
2007).

In a recent study, Wang et al. (2006; hereafter WCHM) estimated a high average
radiative efficiency
of 30\%--35\% for quasars 
at $0.4 < z < 2.1$, implying that most supermassive black holes
are rapidly rotating.
However, the existence of rotating BHs with $\eta\gtrsim 0.18$
is not confirmed by magnetohydronamic (MHD) simulations:
gas loses more angular momentum prior to accretion in an MHD disk
than in a standard thin disk (Gammie \& Shapiro 2004; Shapiro 2005).
Observationally,
Shankar et al. (2008)
present and review evidence for
$\eta\lesssim 0.1$.
In this work we show that the WCHM method is not
independent of quasar lifetimes.  In \S~2 we correct and extend
the WCHM method for determining radiative
efficiencies, in \S~3 we apply it to a sample of Sloan Digital Sky
Survey (SDSS) quasars and in \S~4 we discuss our conclusions.

\section{The Method:}
We assume that quasar light derives only from accretion of matter
onto a black hole, neglecting the effects of BH mergers. Suppose
that mass propagates through a thin accretion disk around an
accreting BH at a rate of $\dot{M}_{acc}$ during some time
interval of $\Delta t$.
The BH mass growth rate is given by
$\dot{M}=(1-\eta)\dot{M}_{acc}$, where $\eta$ is the
radiative efficiency. The quasar radiates at
bolometric luminosity $L$ (Marconi et al. 2004) given by:
\begin{equation}\label{mass_i}
L\Delta t=\eta\dot{M}_{acc}c^{2}\Delta t
=\frac{\eta}{1-\eta}\dot{M}c^{2}\Delta t.
\end{equation}
We can rewrite the above equation as:
\begin{equation}\label{eta_1}
\eta = { L\Delta t \over L\Delta t+\dot{M}c^2\Delta t} \equiv
{\delta\epsilon \over \delta\epsilon + \delta\rho c^2}
\end{equation}
where $\delta\epsilon\equiv L\Delta t/V_{com}$ is the change over
the time $\Delta t$ in the comoving radiative energy density and
$\delta\rho\equiv \dot{M}\Delta t/V_{com}$ is the accompanying
change in the comoving BH mass density, both due solely to this
BH's accretion.

By analogy, an average radiative efficiency can be defined for any
sample of quasars at redshift $z$:
\begin{equation}\label{eta_2}
\bar{\eta}(z)
\equiv\frac{\Delta\varepsilon(z)}{\Delta\varepsilon(z)+\Delta\varrho(z)
c^2}
\end{equation}
where $\Delta\varepsilon(z)$ and $\Delta\varrho(z)$ are the
estimated changes in the \textit{cumulative} radiative energy
density and the \textit{cumulative} mass density of the sample in
the redshift range $(z,z+\Delta z)$.
%
We now define these cumulative densities.

We require the quasar black hole mass function $n(M,z)$, defined
such that $n(M,z)~\Delta M~\Delta z$ is the comoving number
density of black holes with masses in the range ($M,M+\Delta M$)
in the redshift range $(z,z+\Delta z)$. We also require the quasar
bolometric luminosity function $\psi(L,z)$, where
$\psi(L,z)~\Delta L~\Delta z$ is the comoving number density of
black holes with bolometric luminosity in the range ($L,L+\Delta
L$) in the redshift range $(z,z+\Delta z)$.

Over its lifetime $t_q$, a single quasar accretes a mass
$M_{final}$=$\dot{M}_{avg} t_q$ and radiates an energy
$L_{avg} t_q$. We need to express those quantities in
terms of the observables $M_i$ and $L_j$, which are the black
hole's mass and bolometric luminosity at the redshift of
observation, $z$.
We do so by defining correction factors $\bar{f}$ and $\bar{g}$
such that
\begin{equation}
 \bar{f}={1\over N}\sum_{i=1}^{N}{M_i \over M_{final,i}}
=\left<{M_i \over M_{final,i}}\right>
~~{\rm and}~~ \bar{g}={1\over N}\sum_{j=1}^{N}{L_j \over L_{avg,j}}
=\left<{L_j \over L_{avg,j}}\right>
\end{equation}
where the averages are over all $N$ quasars in the sample.

First, consider the mass density that contributes to 
Eq.~\ref{eta_2}. Over
their lifetimes, all black holes observed at redshift $z$ will
accrete a comoving matter density of
\begin{equation}
\sum_{M_i} n(M_i,z) M_{final} \Delta M_i = \sum_{M_i} n(M_i,z) M_i
\Delta M_i/\bar{f}.
\end{equation}
The cumulative, lifetime amount of matter accreted by all black
holes observed at $\geq$$z$ and above is the comoving lifetime
mass density summed over all redshifts $\geq$$z$:
\begin{equation}
\varrho(z)=\sum_{z_k>z} \Delta\varrho(z_k)
=\sum_{z_k>z} \Delta z_k \sum_{M_i} n(M_i,z_k) M_i \Delta
M_i/\bar{f}
\end{equation}
which agrees with Eqs. 2 and 6 of WCHM if $\bar{f}=1$.~Notice that
units of $\Delta\varrho/\Delta z$ are mass per comoving volume per
redshift.

Now consider the radiative energy component of Eq.~\ref{eta_2}.
Over their lifetimes, all black holes observed at redshift $z$
will radiate a comoving energy density of
\begin{equation}
\sum_{L_j} \psi(L_j,z) L_{avg} \bar{t}_q \Delta L_j =
\sum_{L_j} \psi(L_j,z) L_j \bar{t}_q \Delta L_j/\bar{g},
\end{equation}
where $\bar{t}_q$ is the average quasar lifetime.\footnote{
Time periods without accretion are not counted in $\bar{t}_q$.
While $\bar{t}_q$ can be defined as the sum of all time periods
during which an average quasar is actively accreting mass, an acceptable
observational definition might be the time required to accrete,
e.g., 95\% of the quasar's final mass (see Hopkins et al. 2006).}
The cumulative, lifetime amount of energy radiated by all
black holes observed at redshift $z$ and above is the comoving
lifetime energy density summed over all redshifts above
$z$:\footnote{For convenience, in Equations 6 and 8 the redshift
sum appears in front of the mass or luminosity sum for that
redshift bin.  That latter sum is computed in the $k$th redshift
bin, multiplied by $\Delta z_k$ and then added to the mass or
luminosity sum from the ($k$+1)th redshift bin times $\Delta
z_{k+1}$, and so on. The redshift bin size does not matter as long
as $n(M,z_k)$ or $\psi(L_j,z_k)$ does not change considerably
within a bin. For example, if the redshift bin width was halved,
each term in the redshift sum would be half as large but there
would be twice as many terms, yielding the same result.}
%
\begin{equation}
\varepsilon(z)=\sum_{z_k>z} \Delta\varepsilon(z_k)=
\sum_{z_k>z} \Delta z_k \sum_{L_j} \psi(L_j,z_k) L_j \bar{t}_q
\Delta L_j/\bar{g}
\end{equation}
which differs from Eqs. 3 and 5 of WCHM.
Their expression for $\Delta\varepsilon(z)$ is a factor of $\Delta
t_k / \Delta z_k \bar{t}_q$ times the true value above (assuming
$\bar{g}=1$), where $\Delta t_k$ is the cosmological time spanned
by the redshift interval $\Delta z_k$.
The units of $\Delta\varepsilon/\Delta z$ above
are energy per comoving volume per redshift, whereas in Eq. 5 of
WCHM they are energy per comoving volume per (redshift)$^2$, which
is incorrect.

The first quantity of interest, the change in the \textit{cumulative}
comoving mass density of actively accreting black holes over the
redshift range $(z,z+\Delta z)$ --- $\Delta\varrho(z)$
--- is $\Delta z$ times the sum of the masses of all individual
quasar black holes in the sample in that $z$
range, divided by the comoving volume:
\begin{equation}\label{rho_1}
\Delta\varrho(z) = \Delta z~\sum_{M_i} n(M_i,z)~M_i~\Delta
M_i/\bar{f}.
\end{equation}
In other words, the change in the \textit{cumulative} comoving
mass density in the redshift bin $(z,z+\Delta z)$, which is $\Delta\varrho(z)/\Delta z$,
is the same as the comoving mass density in that bin.
Similarly, the second quantity of interest, the change over the
redshift range $(z,z+\Delta z)$ in the cumulative radiative energy
density ever observed from all quasars --- $\Delta\varepsilon(z)$
--- is
\begin{equation}\label{si_1}
\Delta\varepsilon(z)= \Delta z~\sum_{L_j} \psi(L_j,z)~L_j~\Delta
L_j~\bar{t}_q/\bar{g} .
\end{equation}

These expressions for $\Delta\varrho(z)$ and
$\Delta\varepsilon(z)$ can be substituted into Eq.~\ref{eta_2} to
find $\bar{\eta}(z)$ for the sample of quasars under
consideration. Doing so, and grouping $\bar{t}_q$,
$\bar{f}$ and $\bar{g}$ together, we obtain:
\begin{equation}\label{fg}
\bar{\eta}(z) = { \displaystyle\sum_{L_j} \psi(L_j,z)~L_j~\Delta
L_j \over \displaystyle\sum_{L_j} \psi(L_j,z)~L_j~\Delta L_j +
{\bar{g} c^2 \over \bar{f} \bar{t}_q} \displaystyle\sum_{M_i}
n(M_i,z)~M_i~\Delta M_i }
\end{equation}
where the $\Delta z$ in the numerator and denominator have
cancelled out, making the calculation independent of $\Delta
z$ even if $\bar{t}_q$ is longer than the cosmic time interval corresponding to $\Delta z$.
\footnote{For example, consider a population of quasars
observed at a redshift $z_1$, and a population observed at a
redshift $z_2 > z_1$. Those quasars are different objects in
different regions of the universe, but the low-z population might
still be the descendents of the high-z population.  (If we could
watch both regions of the universe over cosmic time, we might see
the high-z population evolve into the low-z population.) In that
case, the masses of the quasars would be counted twice, at
different cosmic times, but the radiative output would also be
counted twice, at those same times.  The estimates of the comoving
mass and radiative energy densities would be systematically off,
but in such a way that the radiative efficiency calculation would
still be accurate.}


Thus, the WCHM method for studying radiative efficiencies
is not independent of the average quasar lifetime $\bar{t}_q$. A
factor of $\bar{t}_q$ enters because both the mass-energy growth
and the radiative energy output of a quasar must be summed over
its entire lifetime (or over the same portion of its lifetime).
The mass-energy sum yields the final mass-energy of the black hole
$M_{tot}c^2$, while the radiative energy sum yields
$L_{avg}\bar{t}_q$.
However, this method remains independent of obscured sources and
is also powerful because it can be implemented for any sample of
quasars regardless of selection effects.  It requires only that
the changes in the cumulative mass-energy and radiative energy
densities be computed using the same objects. Of course, selection
effects {\em will} determine if the resulting
radiative efficiency is relevant for quasars in general.

Also, WCHM in effect assumed $\bar{f}=\bar{g}=1$. We adopt more realistic
estimates by examining Fig. 14 of Springel et al. (2005)
and Fig.~1 of Hopkins et al. (2005), from which we
respectively estimate $\bar{f}\simeq 1$ and $\bar{g}\simeq 10$ for those models.
Adopting $\bar{f}$=1 means that a quasar is observed
when its black hole has accumulated essentially all of its mass, while $\bar{g}$=10
means that the observed bolometric luminosity of a quasar is around 10 times larger
than the average luminosity of a quasar throughout its entire life. Our value of $\bar{g}$ is estimated by
averaging $L_j/L_{ave}$ over the period of time when quasar shows its final burst of activity
(between 1.2 to 1.4 Gyr in Fig.~1 of Hopkins et al. 2005).

If WCHM had assumed $\bar{f}=1$ and $\bar{g}=10$, we estimate that
their (incorrect) calculation would have yielded $\eta\lesssim 0.1$
instead of $\eta\simeq 0.32$. However, Eq. 11 shows that even
with realistic $\bar{f}$ and $\bar{g}$ values the average quasar
radiative efficiency is dependent on the average quasar lifetime.
We now explore the implications of this dependency.
\begin{figure}[ht]
  \epsscale{.6}
    \plotone{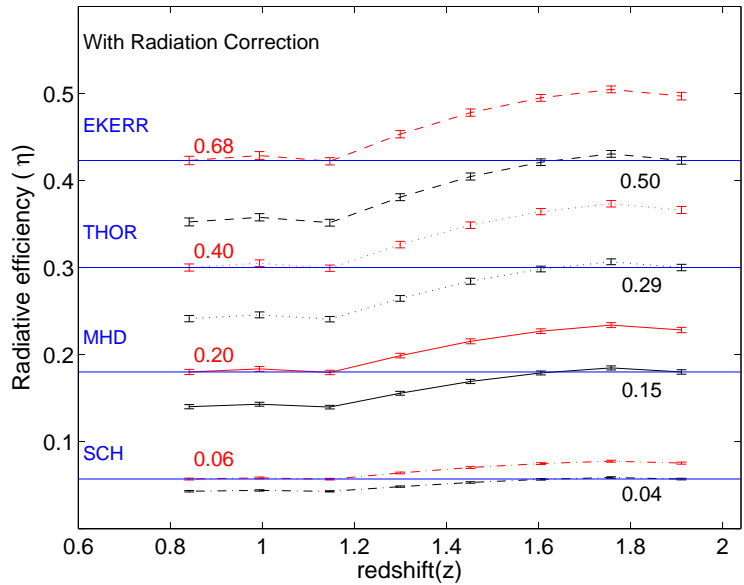}
   \caption{Tracks of putative radiative efficiency $\eta$ vs. redshift z for different
assumed values of the quasar lifetime in Gyr, all assuming
$\bar{f}=\bar{g}=1$. The blue horizontal lines show $\eta$ for the
Extreme-Kerr BH limit (EKERR), the Thorne BH limit (THOR), the
magnetohydrodynamic BH limit (MHD) and the Schwarzschild BH limit
(SCH) respectively. The assumed quasar lifetimes are given next to each
track, on the left for tracks chosen to match a particular radiative efficiency
line at low redshift epochs and on the right for
high redshift epochs. An Extreme-Kerr BH with
$\bar{t}_q\simeq 0.68$ Gyr at $z\simeq0.9$ would violate the maximum possible radiative efficiency
at $z\simeq2$. This inconsistency indicates the existence of a
hidden parameter; namely, Eddington ratio which is considered in Figure \ref{Upsilon}.}\label{1TQ}
\end{figure}

%
\section{Application to SDSS Quasars}
\subsection{Estimation of Black Hole Masses and Bolometric Luminosity}
Based on reverberation mapping studies of local active galaxies,
an empirical scaling relationship has been developed to estimate
black hole masses by (e.g.) Kaspi et al. (2000), Vestergaard
(2002) and McLure et al. (2002). The Sloan Digital Sky Survey
(York et al. 2000) Data Release 3 quasar catalog (Schneider et al.
2005) provides us with a large sample of quasars at redshifts $0.7
< z < 2.1$ which have their \ion{Mg}{2} $\lambda$2800 emission
line redshifted into the SDSS spectral range. We have estimated
black hole masses for 27728 such quasars from the dispersion of
the \ion{Mg}{2} emission line and the continuum luminosity at
$\lambda$=3000 {\AA} and for two scenarios. Scenario A: we assume that
a black hole mass can be estimated from a conventional virial relationship as: $M_{BH}/M_{\odot}=30.5 [\lambda
L_{\lambda}]^{0.5}\sigma_{\rm Mg}^{2}$ where $L_\lambda$ has units
of $10^{44}$ erg s$^{-1}$ and $\sigma_{\rm Mg}$ km s$^{-1}$.
Scenario B: we follow Marconi et al. (2008) in considering
the effect of the radiation pressure of the quasar's radiation on black hole mass
estimates, namely that it reduces the effective gravity on the broad emission
line region, yielding narrower lines at a given mass.
We adopt the BH mass relationship with this radiation correction to be:
$M_{BH}/M_{\odot}=5.75
\bar{\zeta}_\sigma[\lambda L_{\lambda}]^{0.5}\sigma_{\rm
Mg}^{2}+\bar{\xi}_\sigma~[\lambda L_{\lambda}]$ where
$\bar{\zeta}_\sigma=2.4\pm1.5$ and
log$\bar{\xi}_\sigma=6.9\pm0.5$ (Rafiee et al. 2008, in prep.).



Furthermore a Malmquist-like bias has been estimated following \S~4 of Shen et al. (2008).
Our black hole mass estimates have been adjusted downward by a mean bias of $0.4$ dex for scenario A
and $0.11$ dex for scenario B, arising from the steepness of the BH mass function and the scatter in BH mass estimates.

We have matched our sample with that of Richards et al. (2006a), a
homogeneously selected and statistically complete sample of 15343
DR3 quasars with redshifts $z < 5$ drawn from an effective area of
1622~$\textmd{deg}^{2}$.  This procedure yields a subsample of
6704
quasars. The bolometric luminosities of these quasars have been
estimated from $L_{bol}=C_{\lambda}\lambda L_{\lambda}$ with
$C_{\lambda}=5.15$ for $\lambda$=3000 {\AA} following Shen et al.
2008.


Comoving volumes and luminosity distances
have been calculated using a $\Lambda$CDM cosmology with $h=0.71$,
$\Omega_{\Lambda}=0.74$ and $\Omega_{m}=0.26$ (Spergel at al.
2007). Corrections have been made for the limited areal coverage
of the Richards et al. (2006a) sample and for the 5\%
incompleteness of the SDSS at $0.7<z<2.1$. For more details, see
Fig. 8 of Richards et al. (2006a).
\subsection{Radiative Efficiency and Redshift Binning}
For thin accretion disks, the only disks we consider in this paper, $\eta$ varies from $0.057$ for a
Schwarzschild BH (with $a^\ast\equiv Jc/GM^2=0$, where $J$ is the
angular momentum) to $0.42$ for an Extreme-Kerr BH ($a^\ast=1$).
When one takes into account the effect of radiation energy
captured by the BH (Thorne 1974), the radiative efficiency reaches
a maximum of $\simeq 0.30$ ($a^\ast\simeq 0.998$); we refer to
that case as a Thorne BH.
It is more realistic to assume an MHD disk wherein magnetic turbulence
provides a torque to remove angular momentum from the inflowing gas
(Shapiro 2005); in that case, the maximum radiative efficiency is
$\simeq 0.18$ ($a^\ast\simeq 0.938$).\footnote{We relate $\eta$
to $a^\ast$ assuming co-aligned accretion on to rotating BHs.  King,
Pringle \& Hofmann (2008) have pointed out that the effective $\eta(a^\ast)$
will be different for randomly-aligned accretion.
The conversion from $\eta$ to $a^\ast$ will differ
for each combination of co- and randomly-aligned accretion,
but high $\eta$ will always require high $a^\ast$.}

We divide our sample into twelve redshift bins. In each bin we can
compute $\eta$ for any given value of
$\bar{t}_q$. The results are shown in Figure~\ref{1TQ} as tracks
of $\eta(z)$ for eight different values of $\bar{t}_q$ chosen to
match the $\eta$ of a Schwarzschild, MHD, Thorne or Extreme-Kerr
black hole at either the high or low redshift limit of our sample.
Figure~\ref{1TQ} is the corrected version of Figure 2 of WCHM.
Both figures show the evolution of $\eta$ with $z$ for a
flux-limited quasar sample, {\em assuming} constant $\bar{t}_q$.
However, this figure does not give a complete picture since there
is another degree of freedom not being considered;
namely, the Eddington ratios of the quasars. For example, taken at
face value, Figure~\ref{1TQ} suggests that a quasar with a
lifetime of 0.68~Gyr can be powered by a Extreme-Kerr BH at
$z\simeq 0.9$ while the same quasar would violate the maximum
possible radiative efficiency at $z\simeq 2$. Since quasars in our sample have larger values of the
Eddington ratio at $z\simeq 2$ than at $z\simeq 0.8$, the trend in
Figure~\ref{1TQ} might be explained by the importance of the
Eddington ratio rather than the redshift.

%
\begin{figure}[ht]
  \epsscale{0.7}
    \plotone{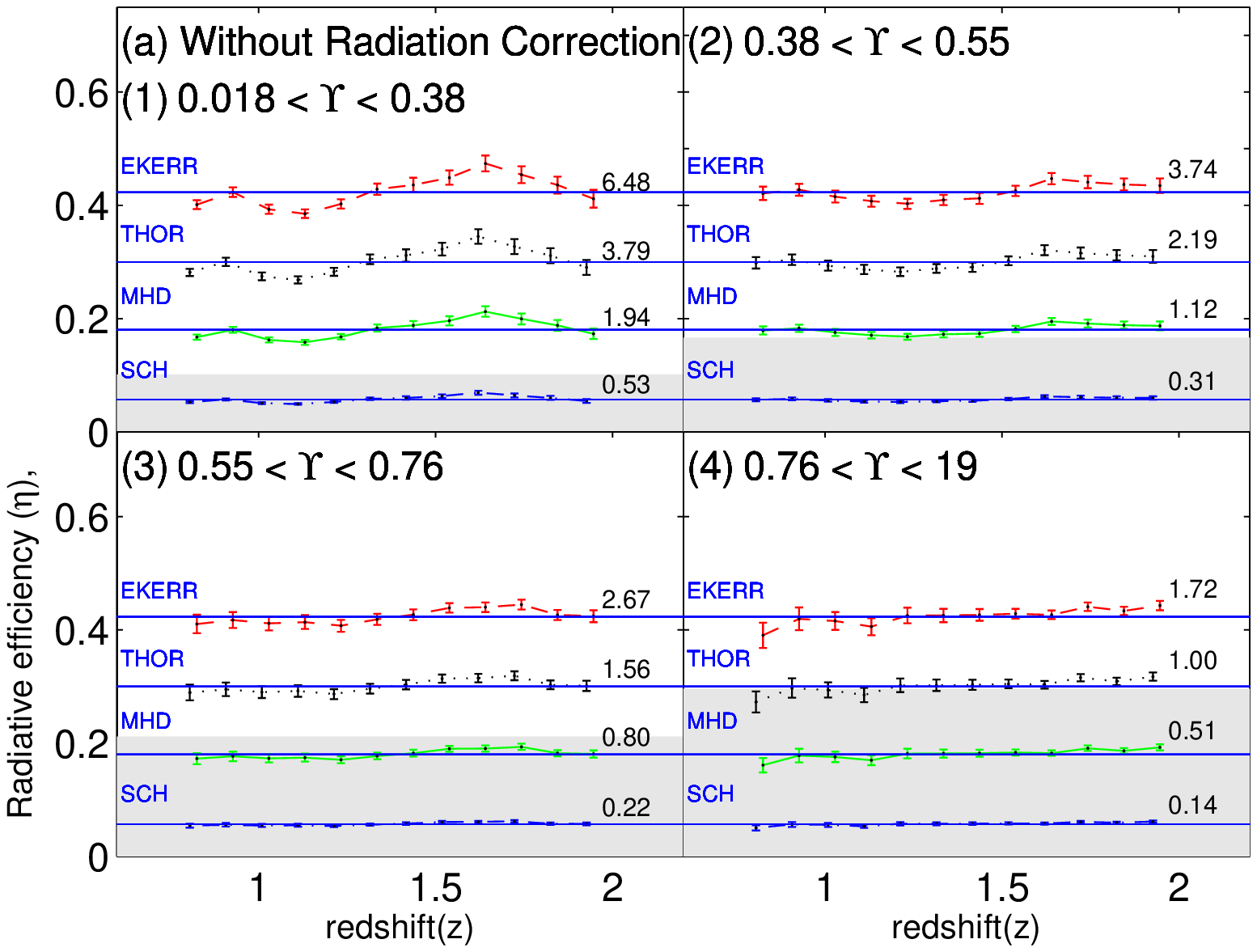}
    \plotone{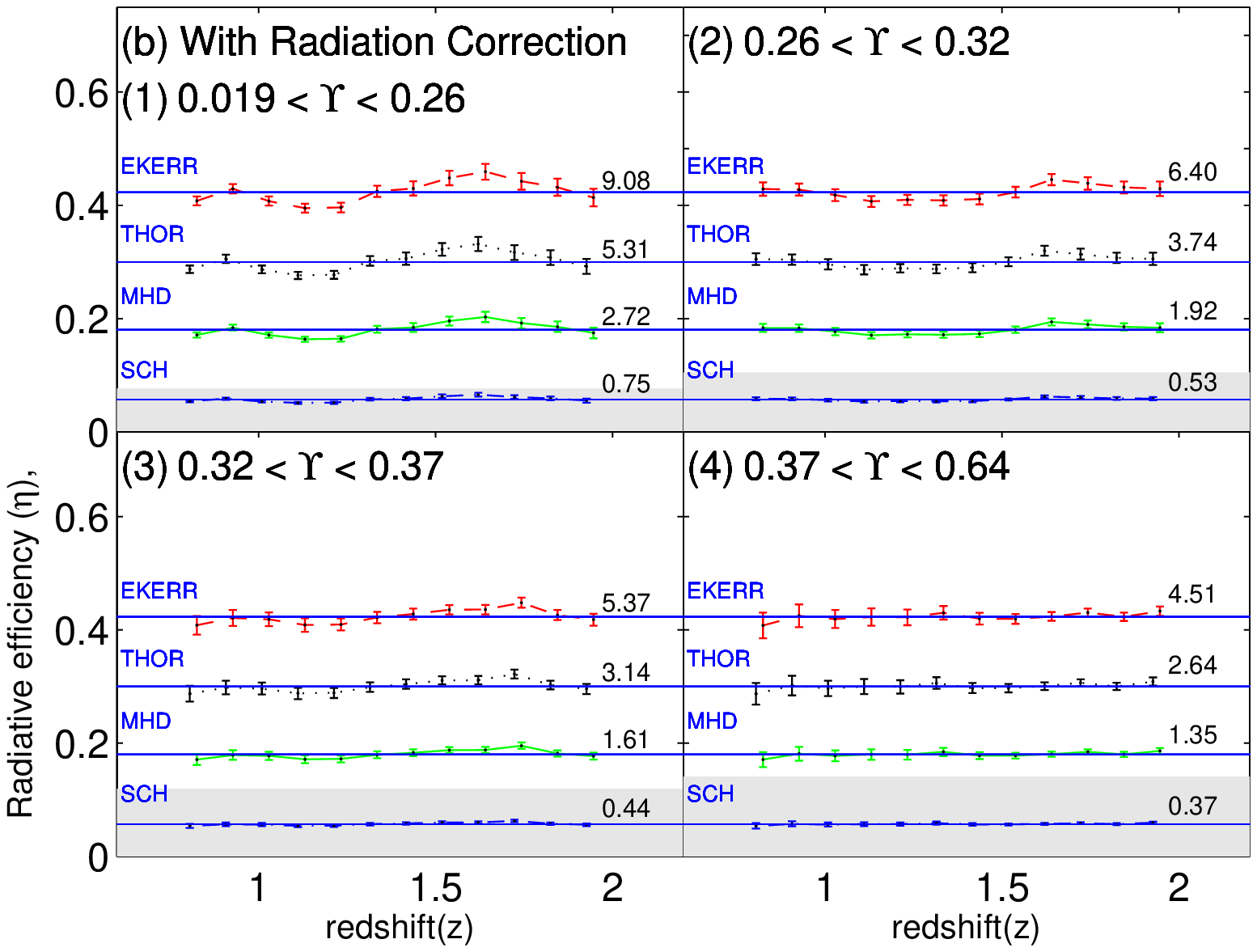}
   \caption{Radiative efficiencies vs. $z$ for different Eddington
ratio bins and constant assumed quasar lifetimes in Gyr (printed
in each panel for each BH type). (a) Scenario A. (b) Scenario B. Lifetimes were chosen so the
resulting $\eta$ would match that of each BH type. We assume
$\bar{f}=1$ and $\bar{g}=10$ --- quasars are observed at their
final mass and ten times their average luminosity.
The shaded areas are the regions corresponding to $\bar{t}_q \lesssim 1$ Gyr,
the possible upper limit for quasar lifetime suggested by Marconi et al. (2004).
}\label{Upsilon}
\end{figure}

\begin{figure}[ht]
    \epsscale{0.7}
    \plotone{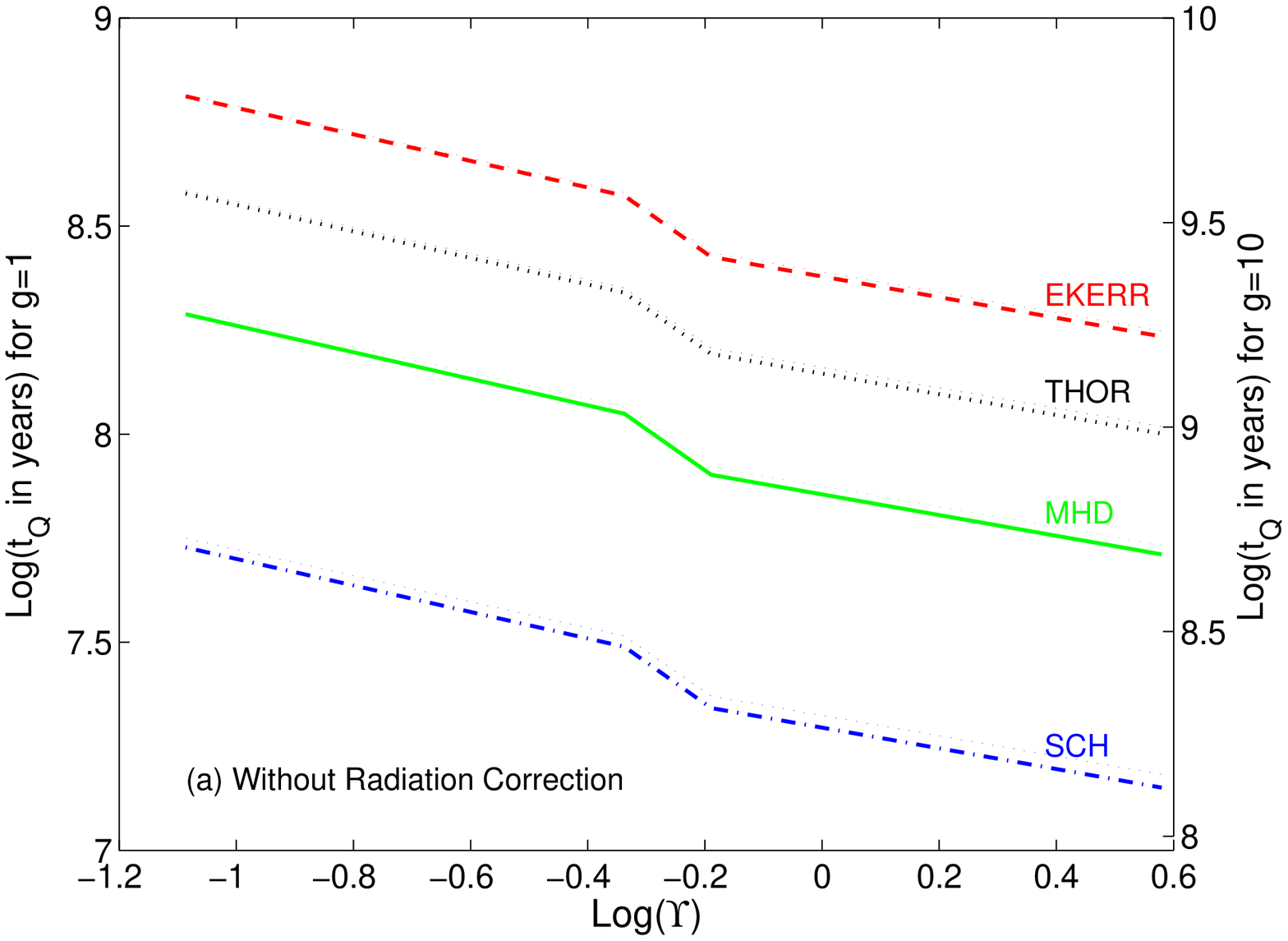}
    \plotone{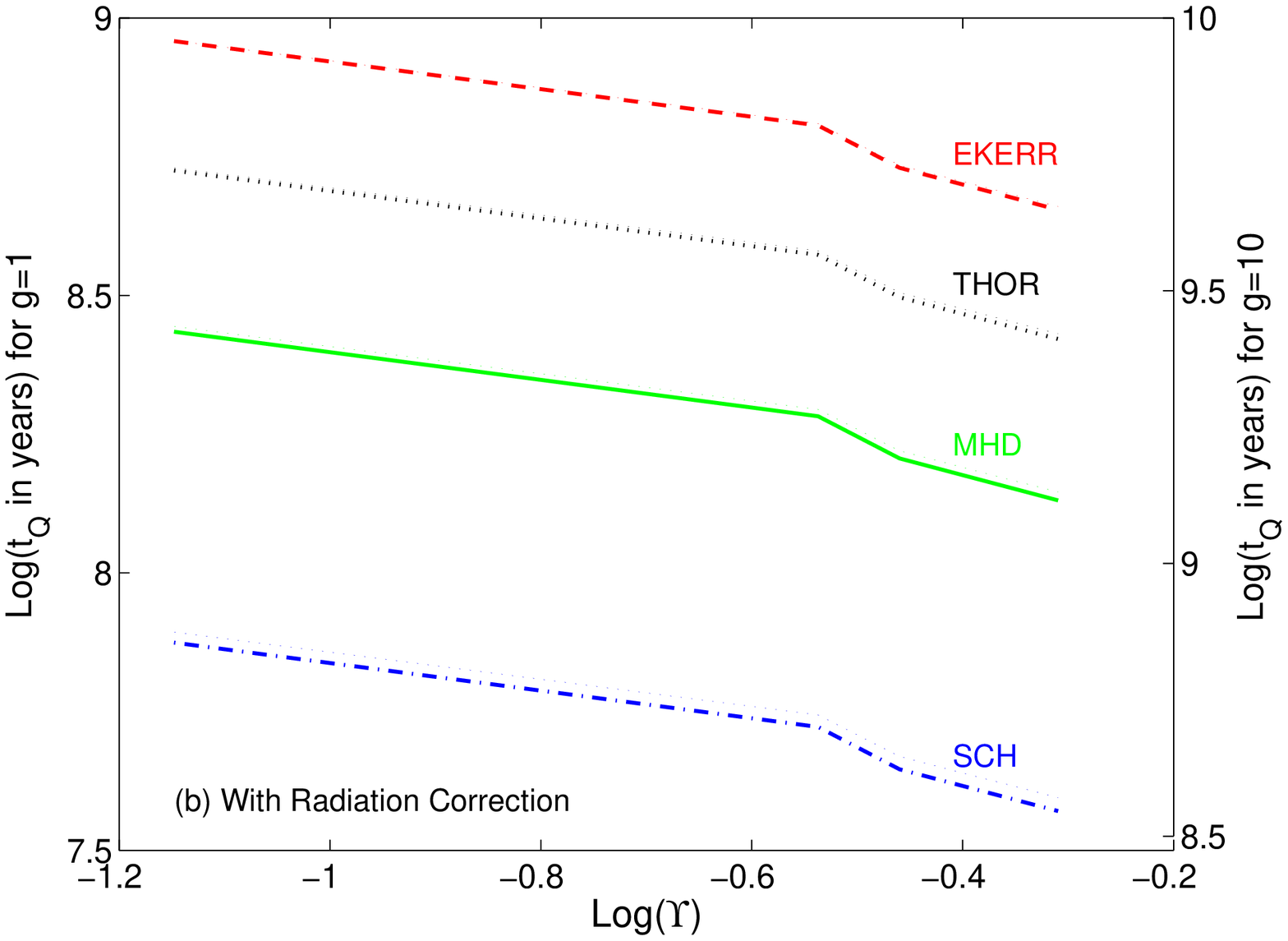}
  \caption{ Quasar lifetime versus Eddington ratio. (a) Scenario A. (b) Scenario B.
  Quasars will lie in the area between the Schwarzschild and Thorne curves
  if they are described by radiatively efficient accretion disks.
  Below the Schwarzschild curves requires radiatively inefficient accretion
  or $\bar{\Upsilon}>\Upsilon_{obs}$,
  while above the MHD curves requires
  $\bar{\Upsilon}<\Upsilon_{obs}$. Left axis labels are for $\bar{f}=\bar{g}=1$;
  right axis labels are for $\bar{g}=10$ and $\bar{f}=1$.
   }\label{tQ_EDR}
\end{figure}

\subsection{Eddington Ratio Binning}
We assume the radiative efficiency may be a function of the
Eddington ratio $\Upsilon\equiv L_{bol}/L_{Edd}$, where
$L_{Edd}=1.26\times10^{38}(M_{BH}/M_{\odot})$ erg~s$^{-1}$.
In that case, the changes in the \textit{cumulative} comoving mass
and energy densities in the Eddington ratio bin
($\Upsilon,\Upsilon+\Delta\Upsilon$) are:
\begin{equation}\label{rho_edd}
\Delta\varrho_\bullet(z,\Upsilon) =\Delta\Upsilon~\Delta
z~\sum_{M_i} n(\Upsilon,M_i,z)~M_i~\Delta M_i/\bar{f}.
\end{equation}
\begin{equation}\label{si_edd}
\Delta\varepsilon_\bullet(z,\Upsilon)= \Delta\Upsilon~\Delta
z~\sum_{L_j} \psi(\Upsilon,L_j,z)~L_j~\Delta
L_j~\bar{t}_q/\bar{g}
\end{equation}
where $n$ and $\psi$ now depend on $\Upsilon$ as well as $z$.
The bullet subscript denotes quantities binned
in $\Upsilon$ as well as $z$.

The average radiative efficiency for a given $\Upsilon$ and $z$ is
\begin{equation}\label{eta_edd}
\bar{\eta}(z,\Upsilon)
=\frac{\Delta\varepsilon_\bullet(z,\Upsilon)}
{\Delta\varepsilon_\bullet(z,\Upsilon)+\Delta\varrho_\bullet(z,\Upsilon)
c^2}.
\end{equation}
For the same sub-sample used in section 3.2 and for twelve
redshift bins, ten mass bins, and ten luminosity bins, the
radiative efficiency has been estimated for four different
Eddington ratio bins (Figure \ref{Upsilon}), each containing one quartile of the objects.
Figure \ref{Upsilon} shows that {\em radiative efficiency is not a
function of redshift but rather of quasar lifetime and Eddington ratio}.

\section{Discussion and Summary}
Determinations of quasar lifetimes, Eddington ratios and radiative
efficiencies are interrelated. Given constraints on (or
assumptions about) quasar lifetimes, the WCHM method can be used
to constrain quasar radiative efficiencies and BH spins. ({\em
Without such constraints, the average quasar $\eta$ cannot be
estimated by this method.}) Conversely, the range of radiative
efficiencies possible for the full range of BH spins can be used
to constrain average quasar lifetimes, as long as luminous quasars
are not powered by radiatively inefficient accretion flows (RIAFs;
see, e.g., Blandford \& Begelman 1999). For example, for the
$\eta=0.065,\Upsilon\simeq 0.4$
model of Shankar et al. (2008), we predict $\bar{t}_q\simeq 140$
million years in our scenario A or $\bar{t}_q\simeq 370$ million years in scenario B,
which could be used as a further test of those models in comparison to others.

Assuming $\bar{f}=1$ and $\bar{g}=10$ (see the end of \S~2), quasar
lifetimes can be constrained according to the Eddington ratio of
the quasar. Lifetimes estimated this way are within a factor of a
few of literature lifetime estimates. For example, for BHs in the
mass range of our sample ($10^8$$<$$M_{BH}$$<$$10^{10} M_\odot$), a
lower limit lifetime of 530 million years can be established for
black holes with $0.02$$<$$\Upsilon$$<$$0.38$ (Panel 1 of Figure
\ref{Upsilon}a, scenario A) or around 750 million years in scenario B for $0.02$$<$$\Upsilon$$<$$0.26$.
This lower limit corresponds to the Schwarzschild
case, since a rotating black hole at the same $\Upsilon$ will
require a longer lifetime to build up its observed mass.
This lower limit lifetime is less than a factor of two lower than the
mean lifetime of one billion years estimated by Marconi et
al.~(2004) for $\Upsilon=0.1$ and $\eta=0.04$ in the same range of
$M_{BH}$. As another example, a luminous quasar powered by a
relatively low-mass black hole --- which may be a typical early
stage in a quasar's evolution --- will have $\Upsilon\gtrsim 0.4$
and can have a
typical lifetime of 140 to 510 million years (Panel 4 of Figure \ref{Upsilon}a).
This range is only a factor of $\sim 3$ larger than the mean
lifetime of $30 - 130$ million years estimated by Yu \& Tremaine
(2002) for luminous quasars, and is consistent with the mean
lifetime of $100-450$ million years estimated by Marconi et al.
(2004) for super-Eddington accretors.

In principle, given constraints on
$\Upsilon\propto\eta\dot{M}/(1-\eta)M_{BH}$, $\bar{f}/\bar{g}$ and
$\bar{t}_q$ for quasar samples, one could estimate the historical
frequency of RIAF episodes in those quasars by plotting quasar
lifetimes versus Eddington ratios.  For example, consider quasars
lying below the Schwarzschild curves in Figure \ref{tQ_EDR} (the
normalization of which is a function of $\bar{f}/\bar{g}$, as seen
by comparing the two axis in either scenario A or B). Above our lower mass limit of $10^8
M_\odot$, such quasars must either have had a RIAF phase in order
to explain their observed masses, or they must have observed
Eddington ratios lower than their historical average:
$\Upsilon_{obs}<\bar{\Upsilon}$. (In the latter case, the quasars
historically would have been located horizontally to the right in
the diagram, lying between the Schwarzschild and Thorne curves at
a value of $\Upsilon=\bar{\Upsilon}$ sufficient to yield the
observed $M_{BH}$ in the observed $\bar{t}_q$.) Conversely,
quasars lying above the Thorne curves in Figure \ref{tQ_EDR}
require $\Upsilon_{obs}>\bar{\Upsilon}$.  A low $\bar{\Upsilon}$
might result if the BH spin does not increase as fast as the BH
mass does, perhaps due to counter-rotating gas accretion phases.
If $\bar{f}=1$ and $\bar{g}=10$ and $\bar{t}_q<1$ Gyr, and if RIAFs
are unimportant, then not many BHs with $\Upsilon < 0.2$ can be
rapidly spinning. On the other hand, if $\bar{f}=1$ and
$\bar{g}=10$ and $\bar{t}_q<140$ Myr, then RIAFs or BH mergers must be
important for quasars regardless of $\Upsilon$, since only
then could the observed masses be reached in the inferred
lifetimes.

What can we conclude if we assume that the Marconi et al. (2008) correction for the effect of radiation pressure
on quasar BH masses is valid, and that $\bar{t}_q \leqslant 1$ Gyr (Martini et al. 2004; Marconi et al. 2004)?
First of all, most quasars can not be rapidly spinning in that case. Second, most of the quasars in our sample
have a radiative efficiency of
$\lesssim 0.14$ consistent with the results of Yu \& Tremaine (2002). This $\eta$ being lower than the MHD
prediction of Shapiro (2005) might be explained by the effects of BH mergers or by the fraction of maximally
spinning BHs being low, at least in our sample.

Alternatively, if one assumes thick disk accretion, where the relation between the BH spin and radiative
efficiency differs from thin disk accretion, then radiatively inefficient accretion becomes more important
even for MHD accretion and spinning BHs, making a lower $\eta$ more plausible.

In conclusion, the Wang et al. (2006) method, despite its advantages, can only estimate the
radiative efficiency of quasars and ultimately the spin of black holes if we know enough about the accretion process
and the evolutionary history of black holes. Better estimates of $\bar{f}$, $\bar{g}$ and $\bar{t}_q$ from ultimately, however, more detailed evolutionary models,
might improve the reliability of the results. Lack of knowledge about the geometry and dynamics of the accretion
disk limits the level of reliability of this method.\\\\

{PBH and AR are supported in part by NSERC.}


%
\end{document}